# A Route Optimization technique for registered and unregistered CN's in NEMO

M. Dinakaran and Dr. P. Balasubramanie

**Abstract**— As the demand of, requesting the Internet without any disturbance by the mobile users of any network is increasing the IETF started working on Network Mobility (NEMO). Maintaining the session of all the nodes in mobile network with its home network and external nodes can be provided by the basic Network Mobility support protocol. It provides mobility at IP level to complete networks, allowing a Mobile Network to change its point of attachment to the Internet, while maintaining the ongoing sessions of the nodes of the network. The Mobile Router (MR) manages the mobility even though the nodes don't know the status of mobility. This article discusses few basic concepts and limitations of NEMO protocol and proposes two ways to optimize the NEMO routing technique for registered and unregistered Correspondent Nodes (CN) of the Mobile Network Node (MNN).

—————— ◆ ——————

## 1 INTRODUCTION

INTERNET access requirement in heterogeneous environments is increasing. The success of cellular communication shows the interest of users in mobility access. These networks are expected to provide not only voice services, and also the data services. IP is the base technology for future networks, which can provide all kind of services with different access modes like fixed and mobile. But IP was not designed for supporting mobility of users and terminals. The IETF has defined some IP-layer protocols that enable terminal mobility in IPv4 and IPv6 [1] networks. But, these protocols do not support the movement of a complete network that moves as a whole and changing its point of attachment to the fixed infrastructure, that is, network mobility. The IETF created a working group: NEMO (Network Mobility), with the aim of extending existing host mobility solutions to enable the movement of networks in IPv6.

Basically IP networks were not designed in terms of supporting for mobility or mobile environments. IP addresses are locators that specify, based on a routing system, how to reach the terminal that is using that address, it can also part of the end-point identifiers of a communication, and upper layers use the identifiers of the peers of a communication to identify it.

The IETF has been working for the problems in terminal mobility; the NEMO group in IETF comes up with IP layer solutions for both IPv4 and IPv6 that enable the movement of terminals without stopping their ongoing sessions. These solutions are even being completed with proposals that improve the efficiency of the base solution, particularly in micro-mobility environments. The issue of terminal mobility has been analyzed recently in [2].

The first step in adaptation of mobile networks is terminal mobility support in IP networks, but there exists also the need of supporting the movement of a complete network that changes its point of attachment to the fixed infrastructure, maintaining the sessions of every device of the network: what is known as network mobility in IP networks. In this case, the mobile network will have at least a router called as Mobile Router (MR) that connects to the fixed infrastructure, and the devices of the mobile network will obtain connectivity to the exterior through this MR. The IP terminal mobility solution does not support the movement of networks, because of that, the IETF NEMO WG [3] was created to specify a solution, at the IP layer, to enable network mobility in IPv6.

Some of the applications, which use the Internet access, are,
(i) Public transportation systems: These systems would let passengers in trains, planes, ships, etc to access the network.
(ii) Personal networks: Electronic devices carried by people, such as PDA's, photo cameras, etc. would connect through a cellular phone acting as the MR of the personal network.
(iii) Vehicular scenarios: Future cars will benefit from having Internet connectivity, not only to enhance safety, but also to provide personal communication, entertainment, and Internet-based services to passengers

The NEMO working group was developed the basic solution to the network mobility problem in IPv6 networks by modifying the IPv6 host mobility solution (MIPv6). But the solution has to be flexible to deal with different mobile networks configurations, in particular, networks containing different subnets and nested mobile networks.

## 2 RELATED WORK

### 2.1 VANET (Vehicular Ad Hoc Networks)

Future cars having Internet connectivity will not only benefit to enhance safety but also to provide personal communication, entertainment, and Internet-based services to passengers through cellular communication networks. When automobiles are near enough, the network traffic can be switched to Vehicular ad hoc network or VANET. We suppose that every vehicle deploys a Mobile Router and has three interfaces: One is ingress interfaces, which connect the node within vehicle (NEMO), next is egress interfaces, which connect Internet, and last is ad hoc interfaces, which connect the neighboring vehicle and set up multi-hop networks. In normal condition, MR can communicate with other MRs through NEMO Basic support protocol and vehicles. Route solution that we offer can transmit and pass VANET in Vehicle-to-Vehicle. Vehicle-to-Internet can be reached through NEMO BSP. Enabling broader communication facilities is an important contribution to the global trend towards ubiquitous communications [7] so; along with technologies of wireless communication, it is possible to install wireless network equipment in vehicles for people to make network connections. So, technologies like NEMO along with VANET can be used for vehicular network since they pose their own purposes [8]. Average frequency of route changes with times the NEMO communicates through VANET.

ROMSGP (receive on more stable group path) [9] will group nodes according to their velocity vectors. If two vehicles were in different groups, the connection between them is considered unstable. In such situation, a penalty will be added to the routing path. Meanwhile, if a node tries to send a packet, it will search it routing table to find next one with less penalty. Additionally, LET (Link Expiration Time) is consider to choose the most stable path i.e. to do a new route discovery before the link being expired. Mobile host in a wireless network may move with certain mobility patterns, such as regular and random movement patterns. Normally, VANET belongs to the regular movement patterns Su et al [10] propose the use of mobility prediction to improve the performance of ad hoc routing [10] with non-random behaviors. In case, if cars are close enough to communicate directly using an ad hoc network a better bandwidth through the infrastructure can be achieved. The reason is that, although the number of hops can be similar, the communication with the infrastructure will typically use a technology with lower bandwidth than the ad hoc network. Also, the ad hoc route will probably result in lower costs. VANET routing [5] can increase route duration time and throughput, and reduce control overhead.

### 2.2 MANEMO

MANEMO[4] is a relatively new and immature concept. The term MANEMO itself can be loosely defined as describing techniques, which combine the properties of Mobile Ad Hoc Networks (MANETs) and the NEMO Basic Support protocol (NEMO BS) to produce solutions, which benefit. The problem of "route optimization in nested nemo networks" is how to construct paths in a dynamic network, and how to route traffic along these paths in an efficient manner. The solution is unique in that it employs classic routing mechanisms, to maintain an ad-hoc network between the mobile routers in the nemo nest.

NEMO-Centric MANEMO (NCM):

NEMO-centric a solution to apply NEMO in MANETs, in which multi-hop communication between a generic MANET node and infrastructure is achieved passing through at least one NEMO Mobile Router running on a different node i.e. If the NEMOs are using NEMO BS to maintain connectivity, packets sent between 2 NEMOs within the nested structure will traverse a highly inefficient route via each of the HAs of the NEMOs that are in the path between the source and the destination NEMOs that results in sub-optimality is known as Pinball Routing (or Multiangular Routing). This is obviously a highly inefficient process and so accordingly, a number of solutions to optimize this situation have been proposed as part of the IETF NEMO Working Group [11] [12]. The concept of combining MANET and NEMO was suggested as one possible solution, it was born from the observation that when the NEMO Mobile Networks converge in the same location to form a nested NEMO structure, this structure itself (locally) is actually a mobile ad-hoc network of NEMO mobile networks. Therefore, local delivery can be best performed between NEMOs in the Nested NEMO structure using a MANET routing protocol (extended to support network prefixes). Although no specific draft proposal was ever submitted to the NEMO WG, the possibility of combining MANET and NEMO in this manner was mentioned in the NEMO WG RO Space Analysis draft [13]

MANET-Centric MANEMO(MCM):

MANET-centric a solution to apply NEMO in MANETs, in which multi-hop communication between a generic MANET node and infrastructure is achieved transparently by means of the MANET routing protocol, whereas NEMO runs on top of it [14] i.e. It is a collection of NEMOs are by default part of an Ad-hoc structure and for them to move away from this structure is the non default case. In this situation it is the MANET protocol that will perform the bulk of the routing and the NEMO protocol that is engaged in the specialized case (vice-versa to the NEMO-Centric scenario). This specialized case occurs when a NEMO has disconnected from the Ad-hoc structure it originated in and therefore uses NEMO BS tunneling to tunnel packets back into the Mobile Ad-hoc Network swarm[15].

The main distinction between a MANET-Centric and a NEMO- Centric MANEMO approach arises when we consider the location of HAs and the Home Networks in general [16]. With NEMO-Centric MANEMO, a HAs role and its subsequent location follows the same model as with NEMO BS, however with MANET-Centric MANEMO it is intended that the Ad-Hoc structure (the MANEMO) is considered the Home Network of each of the NEMOs that belong to it. This distinction represents a big change in the overall conceptual model, but it doesn't massively alter the fundamental role of the HA itself. Essentially the duty of the HA should still be to tunnel packets to and from the MR, the fact that the bulk of the traffic will be sourced from or sent to nodes located on the Home Network shouldn't effect the HAs operation.

## 3 OPERATION OF NEMO

A mobile network (known also as a "network that moves," or NEMO) is defined as a network whose attachment point to the Internet varies with time. Figure 1 depicts an example of a network-mobility scenario. The terminology used by the NEMO group names a router that provides connectivity to the Mobile Network (MN) as a Mobile Router (MR). Devices belonging to the mobile network that obtain connectivity through the MR are called Mobile Network Nodes (MNNs) and they are of different types: Local Fixed Node (LFN) is a node that has no mobility specific software; Local Mobile Node (LMN) is a node that implements the Mobile IP protocol and whose home network is located in the mobile network; and Visiting Mobile Node (VMN) is a node that implements the Mobile IP protocol, has its home network outside the mobile network, and it is visiting the mobile network[7].

The Home Agent (HA) is located in the home network of the mobile network which is a location where the addressing of the mobile network is topologically correct. The Correspondent Node (CN) is a node which sends to or receives a message from MNN. Access Router (AR) is the router in visited network in which MR connected when it is moving out of home network in order to connect with its home network. Care of Address (CoA) is the address given to MR when it is mapped with the AR through the visited network. The HA will refer always the CoA for MR address

When any node located at the Internet, known as a CN, exchanges IP datagram's with a Mobile Network Node, the following operations are involved in the communication. When the MR moves away from the home link and attaches to a new access router (AR), it acquires a Care-of-Address (CoA) from the visited link. As soon as the MR acquires a Care-of Address, it immediately sends a Binding Update to it's HA.

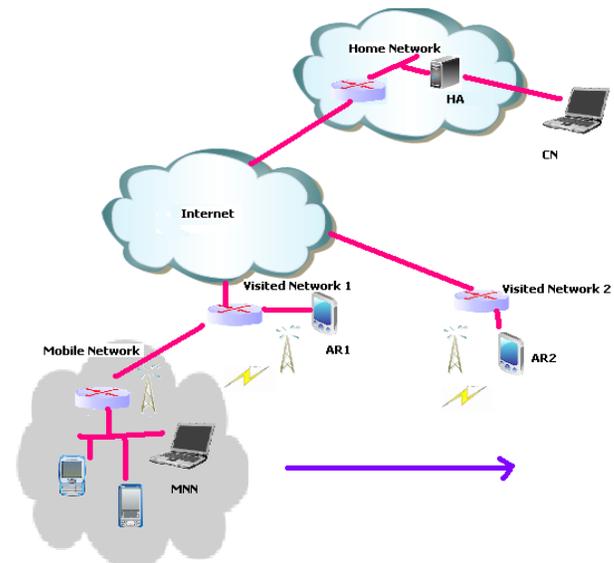

Figure 1 - Network Mobility

When the HA receives this Binding Update, it creates a cache entry binding the MR's Home Address to its Care of Address at the current point of attachment, so that the HA can forward packets meant for nodes in the Mobile Network to the MR. The HA acknowledges the Binding Update by sending a Binding Acknowledgement to the MR. Once the binding process finishes, a bi-directional tunnel is established between the HA and the MR. The tunnel end points are the MR's Care-of Address and the HA's address.

If a packet with source address belonging to the Mobile Network Prefix (MNP) is received from the Mobile Network, the MR reverse-tunnels the packet to the HA. This reverse- tunneling is done by using IP-in-IP encapsulation. The HA decapsulates this packet and forwards it to the CN. When a CN sends a data packet to a node in the Mobile Network, the packet is routed to the HA that currently has the binding for the MR. The HA receives a data packet meant for a node in the Mobile Network; it tunnels the packet to the MR's current CoA.

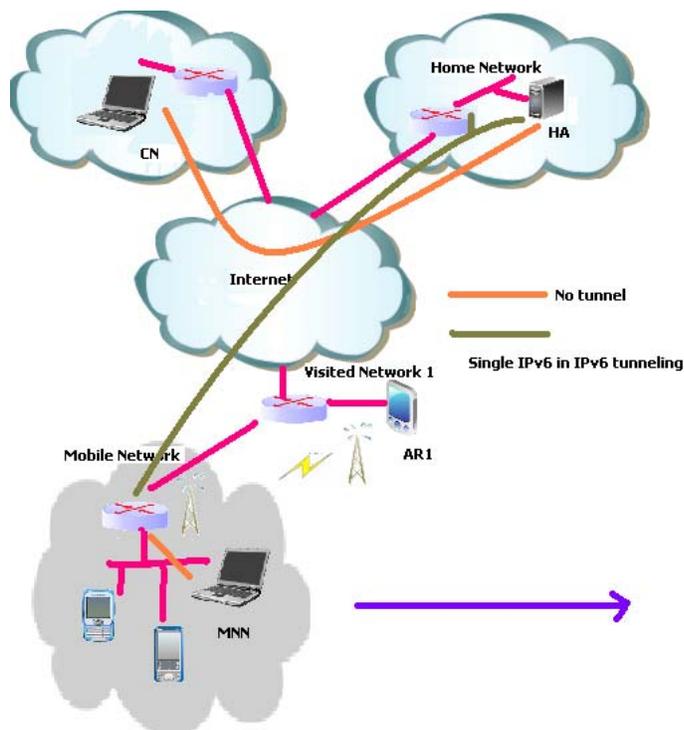

Figure 2 - NEMO Operation

The MR decapsulates the packet and forwards it onto the interface where the Mobile Network is connected. Before decapsulating the tunneled packet, the MR has to check whether the source address on the outer IPv6 header is the Home Agent's address. This check is not necessary if the packet is protected by IPsec in tunnel mode. The MR also has to make sure that the destination address on the inner IPv6 header belongs to a prefix used in the Mobile Network before forwarding the packet to the Mobile Network. If it does not, the MR should drop the packet.

The Mobile Network could include nodes that do not support mobility and nodes that do. A node in the Mobile Network can also be a fixed or a MR. The protocol described here ensures complete transparency of network mobility to the nodes in the Mobile Network. Mobile Nodes that attach to the Mobile Network treat it as a normal IPv6 access network and run the Mobile IPv6 protocol. The MR and the HA can run a routing protocol through the bi-directional tunnel; In this case, the MR need not include prefix information in the Binding Update. Instead, the HA uses the routing protocol updates set up forwarding for the Mobile Network. When the routing protocol is running, the bi-directional tunnel must e treated as a tunnel interface.

The tunnel interface is included in the list of interfaces on which routing protocol is active. The MR should be configured not to send any routing protocol messages on its egress interface when it is away from the home link and connected to a visited link.

Finally, the HA may be configured with static routes to the Mobile Network Prefix via the MR's Home Address. In this case, the routes are set independently of the binding flows and the returning home of a MR. The benefit is that such movement does not induce additional signaling in the form of routing updates in the home network. The drawback is that the routes are present even if the related MR's are not reachable (at home or bound) at a given point of time. The CN transmits an IP data gram destined for MNN-A. This datagram carries as its destination addresses the IPv6 address of MNN-A, which belongs to the MNP of the NEMO.

This IP data gram is routed to the home network of the NEMO, where it is encapsulated inside a new IP datagram by a special node located on the home network of the NEMO, called the HA. The new datagram is sent to the CoA of the MR, with the IP address of the HA as source address. This encapsulation preserves mobility transparency (that is, neither MNNA nor the CN are aware of the mobility of the NEMO) while maintaining the established Internet connections of the MNN. The MR receives the encapsulated IP datagram, removes the outer IPv6 header, and delivers the original datagram to MNN-A. In the opposite direction, the operation is analogous. The MR encapsulates the IP datagram's sent by MNN-A toward it's HA, which then forwards the original datagram toward its destination (that is, the CN).

This encapsulation is required to avoid problems with ingress filtering, because many routers implement security policies that do not allow the forwarding of packets that have a source address that appears topologically incorrect. Additionally, mobile networks can be nested as shown in figure 3. A mobile network is said to be nested when it attaches to another mobile network and obtains connectivity through it.

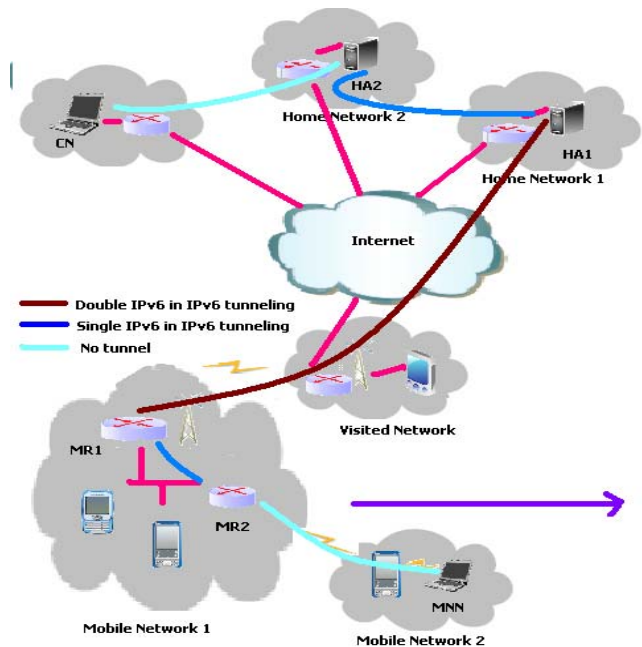

Figure 3 - Nested NEMO

## 4 LIMITATIONS OF NEMO

Given the NEMO Basic Support protocol, all data packets to and from Mobile Network Nodes must go through the HA, even though a shorter path may exist between the MNN and its CN. In addition, with the nesting of MRs, these data packets must go through multiple HA's and several levels of encapsulation, which may be avoided.

This results in various inefficiencies and problems with packet delivery, which can ultimately disrupt all communications to and from the Mobile Network Nodes. The following are the limitations of NEMO Basic Support,

1) Sub-Optimality with NEMO Basic Support: With NEMO Basic Support, all packets sent between a Mobile Network Node (LMN or LFN) and its CN is forwarded through the MRHA tunnel, resulting in a pinball route between the two nodes.
2) Bottleneck in the Home Network: Apart from the increase in packet delay and infrastructure load, forwarding packets through the HA may also lead to either the HA or the Home Link becoming a bottleneck for the aggregated traffic from/to all the MNN. Congestion at home would lead to additional packet delay, or even packet loss. In addition, HA operations such as security check, packet interception, and tunneling might not be as optimized in the HA software as plain packet forwarding. This could further limit the HA capacity for data traffic.
3) Amplified Sub-Optimality in Nested Mobile Networks :By allowing other mobile nodes to join a mobile network, and in particular MR, it is possible to form arbitrary levels of nesting of mobile networks. With such nesting, the use of NEMO Basic Support further amplifies the sub optimality of routing.
4) Security Policy Prohibiting Traffic from Visiting Nodes: NEMO Basic Support requires all traffic from visitors to be tunneled to the MR's HA. This might represent a breach in the security of the Home Network Administrators might thus fear that malicious packets will be routed into the Home Network via the bi-directional tunnel

## 5 PROPOSED ROUTE OPTIMIZATION TECHNIQUES

Basically in NEMO, when data transferred between MNN and CN the bi-directional tunnel is established between the appropriate MR and HA. As data is encapsulated its packet size will increase. If it's nested NEMO the packet size, packet delay and bottleneck in HA will be amplified.

Basically a data transfer or communication between any external node (CN) and MNN can be happened only in two cases they are,
(i) Between MNN and a New CN initiated by CN.
(ii) Between MNN and a known CN initiated either by CN or MNN.

We are proposing route optimization techniques for both the cases.

## 5.1 CASE 1 (MNN AND NEW CN)

When a new CN wants to have communication or data transfer with MNN, it first contacts the agent (HA) of the Home Network.

The usual data transfer flow is CN to HA, HA to MR through AR, finally from MR to MNN. This causes multiple issues as we discussed earlier. To avoid those issues we propose a solution, which can avoid the tunneling and establishes the direct communication between CN to MR for communication or data transfer. As MNN will not have any information or details about the mobility as usual MR is going to work as representative for MNN, because with out MR there is no other way to support mobility to individual nodes in the Mobile Network. But before establishing the direct communication between CN and MR, it must be authorized by HA, so that MR can trust the CN. In this solution we assume that MR will maintains the list of CN's that usually communicates with the MNN's of the Mobile Network. This detail can be maintained as a table like routing table.

When a CN wants to transfer data to a MNN, we propose the following steps to be taken,

1) Once CN initiates data transfer to MNN, it will contact the appropriate HA.
2) The HA will analyze, authorize and forward the data to Care of Address (CoA) of MR after encapsulation.
3) MR will decapsulate the data and checks the CN address.
4) If the CN seems to be new, then it will immediately initiates the binding update request to HA. Then sends the data to MNN.
5) Once the HA acknowledges it, MR will sends the Binding Update to CN directly on behalf of MNN.
6) CN will approve it and add the CoA of MR in its address table.
7) If there is any reply from MNN then MR will forward the same directly to CN.
8) The CN will be added in the MR's CN list. With this table MR will establish direct communication whenever needed.

The steps are illustrated in figure – 4.

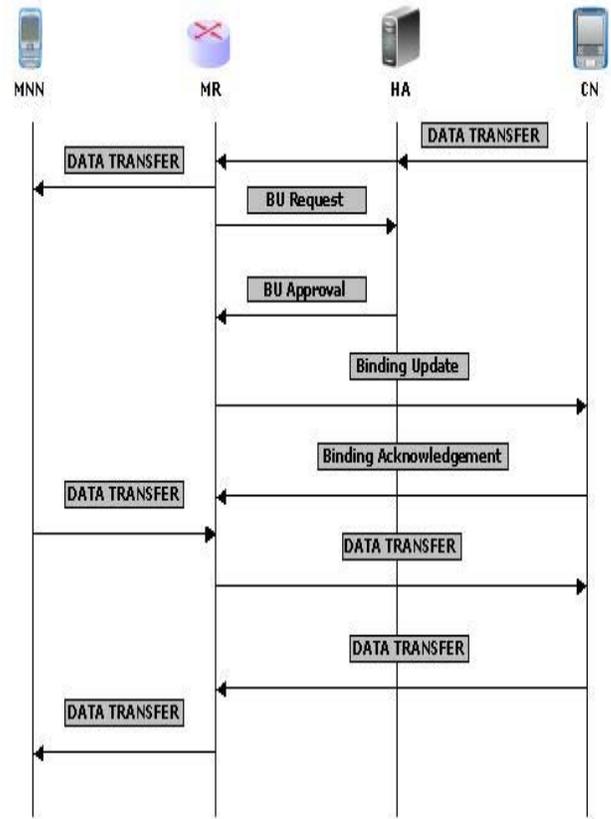

Figure 4 - Direct Communication between MR and CN

## 5.2 CASE 2 (MNN AND EXISTING CN)

When a CN goes for data transfer with and MNN in the mobile network it is automatically added in the MR's table. And those CN will also be adding MR's address in its address table. When CN wants to transfer data to any MNN it will transfer directly to MR, and MR will forward the same to MNN. If MNN initiates data transfer it will send the data to MR, and MR will forward the data to CN. In this case the problem arrives when the MR goes out of the network and gets some other CoA, because CN will not have any information once the MR goes in different access network.

One possible solution for this issues is, whenever the MR changes the access network or it gets new CoA, it will automatically sends the Binding Update (BU) message to all of the CN's registered with it. Once the CN gets the BU from MR, it automatically changes the communication address of the appropriate MNN to this new CoA. With this BU message all the CN's that usually communicates with the Mobile Network will be getting the frequent updates of the CoA from MR [5]. Figure 5 explains the procedure.

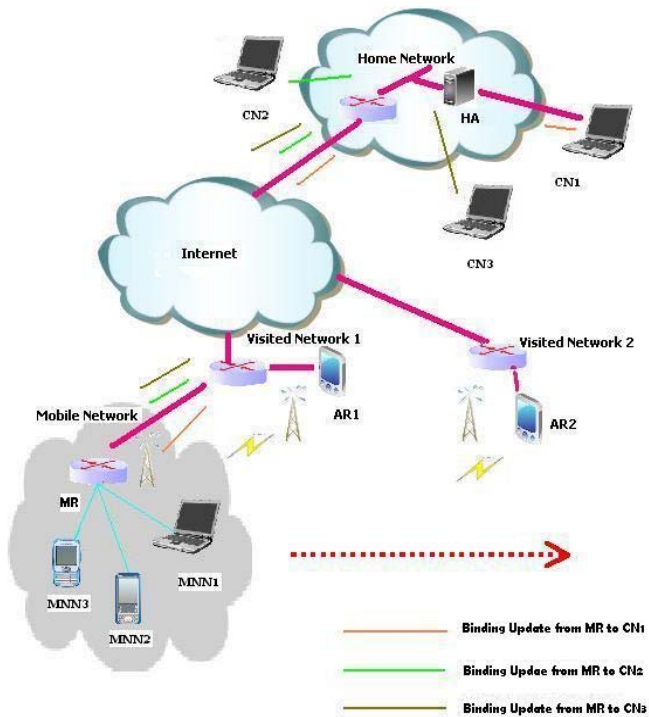

Figure 5 - Binding Update from MR to all CN's of the Network

## 6 CONCLUSION AND FUTURE WORK

We have presented two-route optimization solutions for NEMO in different cases based on Mobile IPv6, which allows the use of the route optimization support for MIPv6 available in CN to provide route optimization for NEMOs. The advantage of this proposal is now where the MNN is disturbed for route optimization or no need of installing any separate program for this technique. Therefore, its adoption would be easier. Further work remains to be done to find a route optimization technique in which we can avoid nested NEMO's.

## ACKNOWLEDGEMENT

We thank Dr. R.S.D. Wahida Banu and Dr.S.P.Shantharajah for giving their valuable suggestions to prepare this article.

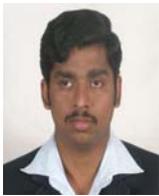
**M. DINAKARAN** has completed his B.Tech (IT) and M.Tech (IT-Networking) in Vellore Institute of Technology, Tamil Nadu and India. After his PG he had 3 years experience in TATA Consultancy Services, India as Asst. System Engineer. He has awarded as TCS Gems during third quarter of 2008. Currently he is working as Asst. Professor in VIT, Vellore and he is pursuing Ph. D in Kongu Engineering College, under guidance of Dr. P. Balasubramanie. He has published 2 articles in International Conferences / Journals.

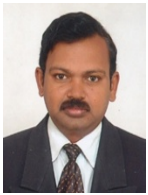
**Dr. P. Balasubramanie** has been awarded Junior Research Fellowship (JRF) by CSIR in the year 1990. He completed his PhD degree in 1996 at Anna University, Chennai. Currently he is a professor in the Department of Computer Science and Engineering in Kongu Engineering College, Perundurai, and Tamilnadu, India. He has guided 7 PhD scholars and guiding 20 scholars Under Anna University. He has published more than 63 articles in International/ National Journals/Conferences. He has authored 6 books with the reputed publishers.